\documentclass[twocolumn]{aastex63}
\pdfoutput=1 
\usepackage{amsmath,amstext,amssymb}
\usepackage[T1]{fontenc}
\usepackage{apjfonts} 
\usepackage{subfigure}
\usepackage{xcolor}
\usepackage{soul}
\usepackage{comment}
\usepackage{graphicx}
\usepackage[normalem]{ulem}
\setstcolor{red}

\graphicspath{{figures/}}

\newcommand{\de}{{\rm d}}

\shorttitle{Standard siren measurement of $H_0$ using O1-O3 GW events}
\shortauthors{A. Palmese et al.}

\begin{document}

\title{
A standard siren measurement of the Hubble constant using gravitational wave events from the first three LIGO/Virgo observing runs and the DESI Legacy Survey}

\author[0000-0002-6011-0530]{A.~Palmese}\thanks{NASA Einstein Fellow}
\affiliation{Department of Physics, University of California Berkeley, 366 LeConte Hall MC 7300, Berkeley, CA, 94720, USA}
\correspondingauthor{Antonella Palmese}
\email{palmese@berkeley.edu}

\author[0000-0003-4383-2969
]{C.~R.~Bom}
\affiliation{Centro Brasileiro de Pesquisas F\'isicas, Rua Dr. Xavier Sigaud 150, 22290-180 Rio de Janeiro, RJ, Brazil}
\affiliation{Centro Federal de Educa\c{c}\~{a}o Tecnol\'{o}gica Celso Suckow da Fonseca, Rodovia M\'{a}rcio Covas, lote J2, quadra J - Itagua\'{i} (Brazil)}

\author{S.~Mucesh}
\affiliation{Department of Physics \& Astronomy, University College London, Gower Street, London, WC1E 6BT, UK}

\author{W.~G.~Hartley}
\affiliation{Department of Physics \& Astronomy, University College London, Gower Street, London, WC1E 6BT, UK}
\affiliation{Department of Astronomy, University of Geneva, ch. d’Ecogia 16, CH-1290 Versoix, Switzerland}

\begin{abstract}
    We present a new constraint on the Hubble constant $H_0$ using a sample of well-localized gravitational wave (GW) events detected during the first three LIGO/Virgo observing runs as dark standard sirens.  In the case of dark standard sirens, a unique host galaxy is not identified, and the redshift information comes from the distribution of potential host galaxies. From the third LIGO/Virgo observing run detections, we add the asymmetric-mass binary black hole GW190412, the high--confidence GW candidates S191204r, S200129m, and S200311bg to the sample of dark standard sirens analyzed. Our sample contains the top 20\% (based on localization) GW events and candidates to date with significant coverage by the Dark Energy Spectroscopic Instrument (DESI) Legacy Survey. We combine the $H_0$ posterior for eight dark siren events, finding $H_0 =  79.8^{+19.1}_{-12.8}~{\rm km~s^{-1}~Mpc^{-1}}$ (68\% Highest Density Interval) for a prior in $H_0$ uniform between $[20,140]~{\rm km~s^{-1}~Mpc^{-1}}$. This result shows that a combination of 8 well-localized dark sirens combined with an appropriate galaxy catalog is able to provide an $H_0$ constraint that is competitive ($\sim 20\%$ versus 18\% precision) with a single bright standard siren analysis (i.e. assuming the electromagnetic counterpart) using GW170817. When combining the posterior with that from GW170817, we obtain $H_0 =  72.77^{+11.0}_{-7.55}~{\rm km~s^{-1}~Mpc^{-1}}$. This result is broadly consistent with recent $H_0$ estimates from both the Cosmic Microwave Background and Supernovae.
\end{abstract}


\section{Introduction}

Since the first detection of gravitational waves (GW) in 2015 \citep{GW150914}, the astrophysics community has significantly intensified its efforts in the field of multi-messenger astronomy. Thanks to these remarkable efforts, it was possible to identify the first electromagnetic counterpart to a GW event \citep{MMApaper,arcavi,Coulter1556,lipunov,tanvir,marcelle17,valenti}, GW170817 \citep{ligobns}, the first binary neutron star merger to be detected in gravitational waves. Multi-messenger observations can be used as powerful probes of cosmology and fundamental physics: they can be used to place constraints on gravity theories and to test General Relativity (e.g. \citealt{Yunes_2013,Baker_2017,Ezquiaga_2017,carson2020testing,palmese20}), to measure the equation of state of neutron stars, and even to understand whether compact object mergers could be made of primordial black holes \citep{2018JCAP...09..039S,tsai2020gw170817}. Another interesting application of multi messenger observations is that of ``standard sirens'', first proposed in \citet{schutz}: given a gravitational wave detection, it is possible to measure a luminosity distance for the event, and if that is combined with the redshift of the galaxy that hosted the merger, gravitational wave events can be used to probe the distance-redshift relation. Since the relation is sensitive to cosmological parameters, GW events can be used to infer those parameters, and in particular the Hubble constant $H_0$ for nearby events.

The standard siren methodology can be applied both to events with a counterpart (``bright standard sirens''), for which a unique host galaxy can potentially be identified, and to events without a counterpart (``dark'' or ``statistical'' standard sirens). In both cases, the redshift information comes from galaxies, whether that is a unique galaxy or a sample of potential hosts. The first bright standard siren measurement of the Hubble constant has been derived in \citet{2017Natur.551...85A} for GW170817, which is also the only event so far with a confident counterpart association with a unique host galaxy (e.g. \citealt{Blanchard_2017,palmese}). Several works \citep{chen2020standard,gayathri2020standard,mukherjee2020standard} have used a candidate counterpart for another event, GW190521, to derive bright standard siren measurements. However, the counterpart association to the GW event cannot be made with confidence \citep{ashton,Palmese_2021}.

Dark standard siren measurements have been produced for a larger number of events, including GW170817 \citep{fishbach}, the binary black hole mergers GW170814 and GW190814 (\citealt*{darksiren1}; \citealt{palmese20_sts,190814_paper}), and a larger ensemble if events from the first two LIGO/Virgo observing runs, O2 \citep{LVC_O2_StS}, and the first half of the third observing run, O3a \citep{finke2021cosmology}. Dark standard sirens require knowledge of the position and redshift of the ensemble of potential host galaxies, over which one needs to marginalize, and are therefore expected to lead to less precise results than bright standard sirens on a single event basis. On the other hand, GW binary black holes, and more in general events without counterpart, currently outnumber those with a counterpart by at least a factor of 10 \citep{gwtc2}. We therefore take the approach of combining dark sirens with the much rarer bright sirens to derive standard siren constraints on cosmological parameters.

New measurements of the Hubble constant are valuable to gain insight on the so-called ``Hubble constant tension'' (e.g. \citealt{Riess_2019,planck18,Freedman_2019,riess2021}). This tension arises from the significant discrepancy between different measurements of $H_0$, namely those that rely on measurements from the Cosmic Microwave Background \citep{planck18}, and those using Supernovae (SN) with a local distance ladder \citep{riess2021}. New, independent measurements of $H_0$ could help clarifying where the tension is arising from (e.g. \citealt{verde}), and GW standard sirens offer such possibility. In particular, dark standard sirens for which higher order modes can be measured (such as one of the events we use in this work) are particularly promising in providing more precise sky localizations and distances, potentially reaching a few per cent uncertanty in $H_0$ in 5 years \citep{Borhanian_2020}.

In this work, we present a new measurement of the Hubble constant using 8 dark sirens, combined with GW170817 using the counterpart. We decide to focus on the best localized events whose high probability volume is also covered by a complete galaxy catalog. These are in fact the most important characteristics to derive more precise constraints on the Hubble constant. First, better localized events need to be marginalized over a smaller number of galaxies than those that encompass large volumes, so that they are more likely to provide better constraints. Better localized events also tend to be less distant, which means that a standard siren analysis would mostly be sensitive to the Hubble constant rather than other cosmological parameters. Lastly, the galaxy catalog coverage is fundamental. 

\citet{LVC_O2_StS} find that the improvement on the Hubble constant measurement coming from dark sirens compared to a GW170817--only analysis is marginal, and this can be attributed to the lack of complete galaxy catalogs, along with the lack of well-localized events in the first observing runs. Likewise, \citet{palmese20_sts} find that the addition of only two well--localized dark sirens with uniform and deep coverage of galaxy catalogs can provide significant improvement to GW170817. First, we improve on the results in \citet{palmese20_sts} by adding new events from O3. We also improve on the analysis in \citet{finke2021cosmology} by considering a more suitable galaxy catalog. The catalog used there is a compilation of galaxies from different surveys up to 2014, and it is complete out to a distance of $\sim 37$ Mpc \citep{glade}, but largely incomplete at the distances of the dark sirens (all at $>200$ Mpc). We use galaxy data from state-of-the-art dark energy experiments, the Dark Energy Spectroscopic Instrument (DESI; \citealt{desi_collaboration_desi_2016}) imaging surveys and the Dark Energy Survey (DES; \citealt{descollaboration}), both completed in 2019. Data from these surveys have been augmented with infrared bands and spectroscopic data, and processed to produce the Legacy Survey catalogs. We use photometric redshifts from \citet{zhou2020clustering}, and also derive alternative redshifts through a method that simultaneously constrains redshift and luminosity of galaxies using random forests \citep{galpro}. These catalogs are used to derive an Hubble constant posterior using the LIGO/Virgo asymmetric mass binary black hole (BBH) merger GW190412 \citep{GW190412}, and the high confidence binary black hole candidates S191204r \citep{191204}, S200129m \citep{200129}, and S200311bg \citep{200311}.

The paper is structured as follows. We describe the data used in \S\ref{data} and our methods in \S\ref{method}. Results and discussion are presented in \S\ref{results}, followed by the conclusions in \S\ref{conclusions}. We assume a flat $\Lambda$CDM cosmology with $\Omega_m =0.3$ and $H_0$ values in the $20-140~{\rm km~s^{-1}~Mpc^{-1}}$ range. When not otherwise stated, quoted error bars represent the 68\%\ credible interval (CI).

\section{Data}\label{data}

\begin{figure*}
\centering
\includegraphics[width=1\linewidth]{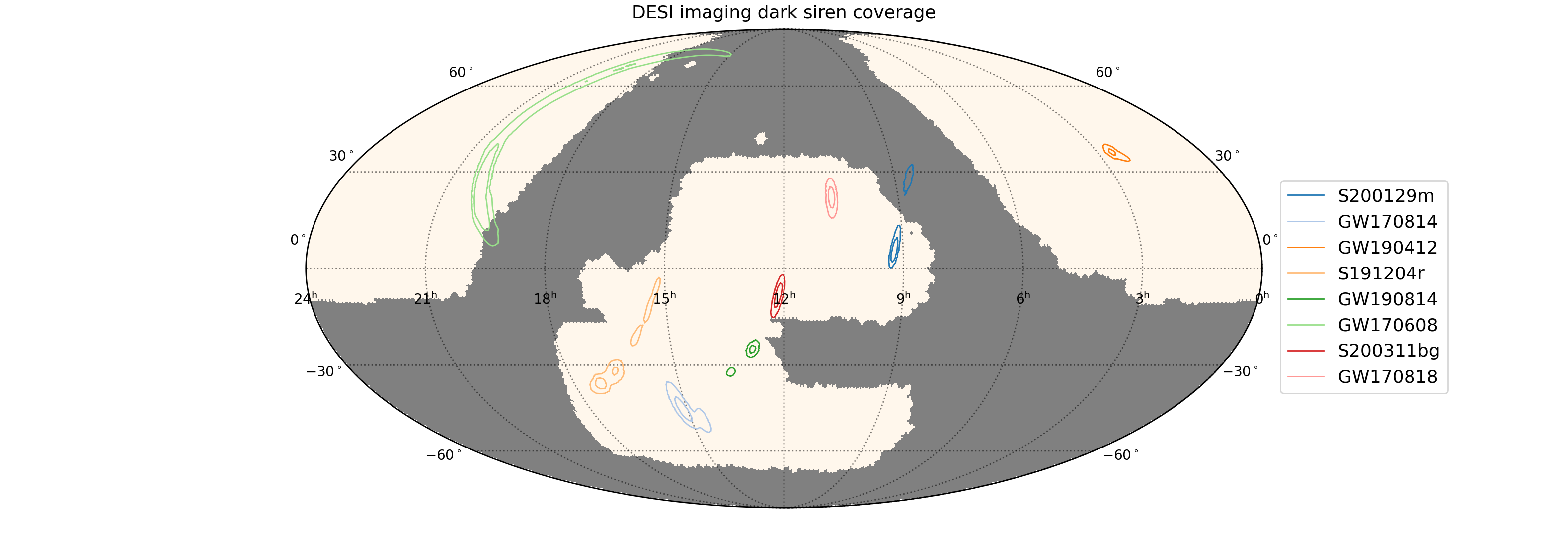}
\caption{LIGO/Virgo GW dark standard sirens analyzed in this paper, where the shaded regions represent the 90\% CI localization from the sky maps. The shaded regions are those that are not covered by the DESI Legacy Survey DR9 catalogs used in this work.}
\label{fig:data}
\end{figure*}

\subsection{The LIGO/Virgo GW data}\label{sec:GWdata}

We select the top 20\% gravitational wave events and high confidence candidates based on their 90\% CI comoving volume. We then select those with $>70\%$ of their probability covered by the DESI imaging.

The GW map we use for GW190412 is one of the two fiducial maps presented in \citet{gwtc2}. We choose to use the one obtained from fitting the effective-one-body (EOB) family of waveform models to the GW signal. These models also include a post-Newtonian (PN) prescription and numerical relativity (NR) information in the inspiral-merger-ringdown description, and the effect of higher-order multipoles for precessing binaries, so they are referred to as SEOBNRv4PHM \citep{2014PhRvD..89h4006P,Ossokine_2020}. The event is localized to within 12 deg$^2$ at 90\% CI, and has a luminosity distance of $740^{+120}_{-130}$ Mpc after marginalization over all other parameters. With a 90\% CI comoving volume of $4 \times 10^{-4}$ Gpc$^3$, GW190412 is the second-to-best dark standard siren to date, only second to GW190814.
At the time of writing, the three candidate events we consider had not been confirmed as gravitational wave events by the LIGO/Virgo Collaboration yet, but their False Alarm Rate (FAR) is so small (see Table \ref{tab:events}), and their classification as BBH is confident ($>99\%$), that they are very likely to be real GW events of astrophysical origin. We note that their FAR is estimated differently from the other events considered here, since it results from the low-latency analyses rather than from the more accurate offline studies. These candidates have all been confirmed as gravitational wave events at the time this work was ready for submission \citet{gwtc3}. Their 90\% comoving volume is comparable to that of a previously analyzed dark standard siren, GW170814, and it is of the order of $\sim 10^{-3}$ Gpc$^3$.

\begin{table*}
\centering
\begin{tabular}{cccccc}
Event & $d_L$ [Mpc] & A [deg$^2$] & V [Gpc$^3$] & FAR  & Reference\\
\hline
GW170608 & $320^{+120}_{-110}$ & 392 & $3 \times 10^{-3}$ &$<1$ per $ 10^5$ yr &  \citet{gwtc1} \\
GW170814 & $540^{+130}_{-210}$ & 62 & $2 \times 10^{-3}$ &$<1$ per $ 10^4-10^7$ yr &  \citet{gw170814}\\
GW170818 & $1060^{+420}_{-380}$ & 39 & $7 \times 10^{-3}$ &$<1$ per $ 10^5$ yr &  \citet{gwtc1}\\
GW190412 & $740^{+120}_{-130}$ & 12 & $4 \times 10^{-4}$  &$<1$ per $ 10^3-10^5$ yr &  \citet{GW190412}\\ 
GW190814 & $241^{+26}_{-26}$ & 19 & $3 \times 10^{-5}$ &$<1$ per $ 10^4-10^7$ yr &  \citet{190814_paper}\\
S191204r & $678^{+149}_{-149}$ & 103 & $6 \times 10^{-3}$ &$<1$ per $ 10^{17}$ yr &  \citet{191204}\\ 
S200129m & $755^{+194}_{-194}$ & 41 & $2 \times 10^{-3}$ & $<1$ per $ 10^{23}$ yr  &  \citet{200129}\\
S200311bg & $1115^{+175}_{-175}$ & 34 & $5 \times 10^{-3}$ & $<1$ per $ 10^{17}$ yr  &  \citet{200311}\\

\hline

\end{tabular}
\caption{Luminosity distance, 90\% CI area and volume, and False Alarm Rate (FAR) of gravitational wave events and candidates used in this analysis. We also report the reference paper or GCN that reports to the sky map used for each event. Where a range of FAR is provided, this is because multiple FAR estimates are available from multiple search algorithms. The FAR reported for the candidates is different from the confirmed events as it is estimated from the online analysis. These candidates have all recently been confirmed as gravitational wave events in \citet{gwtc3}.
} 
\label{tab:events}
\end{table*}

\subsection{The DESI Imaging data}\label{sec:desi}

The DESI collaboration observed a large fraction of the sky ($\sim 14,000$ sq deg) in $grz$ bands out to $\sim2$ mags deeper than SDSS, in order to select their spectroscopic targets from a photometric sample which is large and deep enough to contain a dense galaxy and quasar sample. An area of $\sim 9000$ sq deg of the North and South Galactic Caps (NGC and SGC) up to dec$<32$ deg was imaged in $grz$ by the Dark Energy Camera (DECam; \citealt{flaugher}) mounted on the Mayall 4m telescope at the Cerro-Tololo Inter-American Observatory. These observations was part of the The Dark Energy Camera Legacy Survey (DECaLS). The remaining area of the NGC was imaged in $g$ and $r$ by the Bok 90 inch (as part of BASS, the Beijing–Arizona Sky Survey; \citealt{bass}), and in $z$ by the Mosaic-3 on the Mayall 4m telescope at Kitt Peak National Observatory (as part of MzLS, the Mayall z-band Legacy Survey).

The optical bands are supplemented with IR data from the Wide-field Infrared Survey Explorer (WISE; \citealt{Wright_2010}). Source detection and photometry, including forced photometry on the unWISE $W1$, $W2$, $W3$, and $W4$ images \citep{unwise1,unwise2,unwise3} at 3.4 and 4.6 $\mu$m respectively, is performed with the software package \textsc{The Tractor} \citep{tractor}. The $5\sigma$ depth of the survey is 24.0,23.4 and 22.5 in $grz$ respectively, for the AB mag of a fiducial galaxy with an exponential light profile and half--light radius 0.45 arcsec. Survey and calibration have been performed carefully to ensure a maximally uniform source sample despite the different instruments used. Survey and photometry are described in more detail in \citet{desiimaging}.

We use photometry and photometric redshifts from Data Release (DR) 9, which covers more than $16,000$ sq deg thanks to imaging available outside of the nominal DESI imaging footprint. We select galaxies by requiring \texttt{PSF$\neq 1$}. Since this cut alone provides a highly star-contaminated galaxy sample, we follow \citet{Ruiz-Macias} and apply a further cut using GAIA data. We treat objects as galaxies if either of these conditions apply:
\begin{itemize}
    \item they are not in GAIA;
\item they are in GAIA but have $(G_{\rm GAIA} - r_{\rm raw}) > 0.6$, where $(G_{\rm GAIA}$ is the G band GAIA magnitude, and $r_{\rm raw}$ is the DESI Imaging $r$ band magnitude uncorrected for the Galaxy's dust extinction.
\end{itemize}
Moreover, we correct the photometry for Milky Way dust extinction using the correction factors provided in the Legacy Survey release, which are based on the dust maps of \citet{1998ApJ...500..525S}, and remove duplicated sources in the regions where the north and south surveys overlap, keeping the DECaLS sources where available.

\subsubsection{Redshifts truth table}\label{sec:spec}

As a truth table for training, validation, and testing of redshifts and luminosities of the photometric sample, we use the sample compiled by \citet{zhou2020clustering}, which comprises of spectroscopic redshifts and sophisticated multi--band photo--$z$'s from several galaxy surveys. It includes data from the 2-degree Field Lensing Survey \citep{2df}, the AGN and Galaxy Evolution Survey (AGES; \citealt{ages}), the COSMOS2015 catalog \citep{Laigle_2016}, DEEP2 \citep{deep2}, the Galaxy and Mass Assembly (GAMA;\citealt{gama}, the OzDES \citep{childress_ozdes_2017}, SDSS DR14 \citep{abolfathi_fourteenth_2018} -- including the Main Galaxy Sample \citep{strauss_spectroscopic_2002}, the Baryonic Oscillation Spectroscopic Survey (BOSS; \citealt{dawson_baryon_2013}) and the extended BOSS (eBOS; \citealt{dawson_sdss-iv_2016}) -- the VIMOS Public Extragalactic Redshift Survey (VIPERS; \citealt{scodeggio_vimos_2018}), the VIMOS VLT Deep Survey (VVDS; \citealt{le_fevre_vimos_2013}), and the WiggleZ Dark Energy Survey \citep{parkinson_wigglez_2012}. The objects from these catalogs are selected based on their quality flags as described in \citet{zhou2020clustering}, and matched to DESI imaging sources to create the final truth table, that comprises of more than 2.5 million objects.

The training sample that we use for \textsc{GALPRO} is the same as the one defined in \citet{zhou2020clustering}: the truth table is downsampled in order to reduce the effect of spurious peaks in the redshift distribution of the evaluation sample, due to the selection function of the most numerous samples in the truth table (namely the SDSS Main Galaxy Sample, BOSS, WiggleZ and GAMA). If a training sample presents sharp peaks in the redshift or color distribution because of a specific color or magnitude selection in the truth table, these regions are overrepresented compared to others, and it is likely that the random forest would oversample from the redshifts and colors around these peaks, causing systematic biases.
This selection process is particularly important for our work, since we want to avoid introducing spurious peaks in the $dN/dz$ that could result in spurious peaks in the Hubble constant posterior. The training sample contains $\sim 950,000$ objects with a redshift distribution that is roughly flat over the redshift range of interest here ($z<0.4$). The remaining truth redshifts are used for testing the photo--$z$ performance and measure the systematic bias.
We compute rest-frame $r$-band luminosities ($\textrm{L}^{\textrm{rest}}_{r}$) using the method described in \citet{rudnick}. 

\begin{figure*}
\centering
\includegraphics[width=0.5\linewidth]{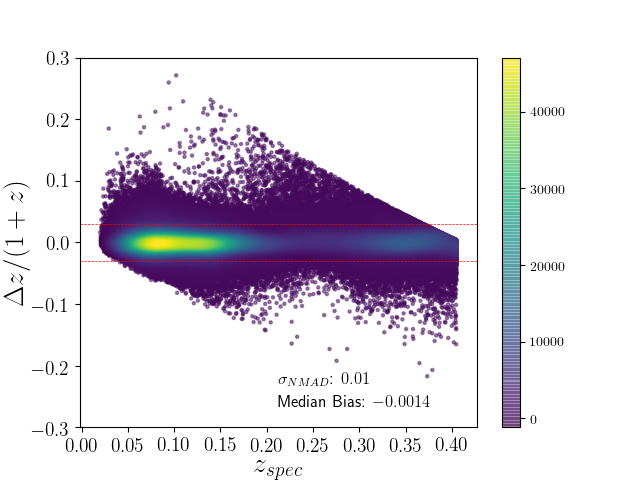}
\includegraphics[width=0.4\linewidth]{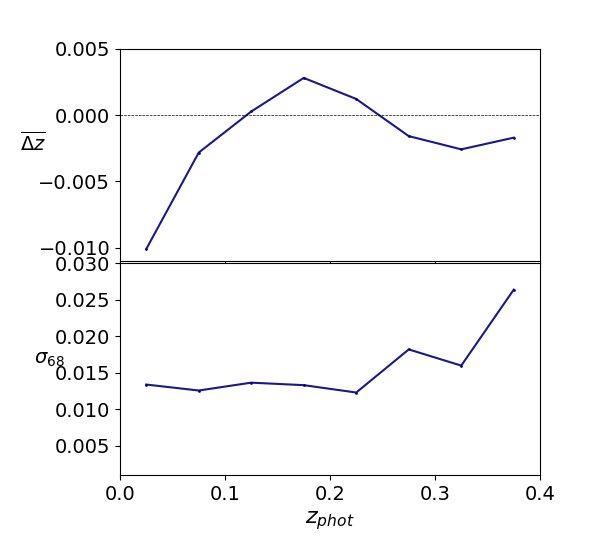}

\caption{Photometric redshift quality assessment plots using a validation sample from the available spectroscopic data overlapping with the DESI Legacy Survey. \emph{Left:} density
plot of galaxies in the validation sample, showing the deviation of the point estimate photo-$z$ $z_{\rm phot}$ from the spectroscopic redshift $z_{\rm spec}$ (where $\Delta z= z_{\rm spec}- z_{\rm phot}$) as a function of spectroscopic redshift. The red dashed lines define the 2$\sigma_{68}$ region.  
\emph{Right:} mean value and scatter $\sigma_{68}$ of the bias distribution $\Delta z$ in bins of photometric redshift.}
\label{fig:data2}
\end{figure*}

\subsubsection{Legacy survey photo-$z$'s}\label{sec:LSphz}

Photometric redshifts have been computed by \citet{zhou2020clustering} for the entire Legacy Survey using Random Forests (RF). Their input features for the RF are $r$--band magnitude, $g-r$, $r-z$, $z-W1$ and $W1-W2$ colors. In addition, they provide three morphological parameters following \citet{Soo_2017}, significantly improving the scatter and outlier fraction of the testing sample. These parameters are the half-light radius, the ratio between semi-minor and semi-major axes, and a ``model weight'' that quantifies how well a galaxy is fit by an exponential light profile versus a de Vaucouleurs profile.

When using these photo--$z$'s, we follow the findings of \citet{zhou2020clustering} and cut the sample at $z<21$ to ensure that the photo-$z$ bias is under control, and at photo-$z$ errors $<0.1$ to remove catastrophic outliers. Here we present the performance of these photo-$z$'s on the redshifts from the truth table that were not used by \citet{zhou2020clustering} as part of their training sample for the redshift range of interest here. This is necessary for our work since the analyses of \citet{zhou2020clustering} focus on a higher--$z$ sample of Luminous Red Galaxies, while we are interested in a lower redshift sample of generic galaxies. From the photo-$z$ catalog we select galaxies with a spectroscopic measurement from the compilation of \citet{zhou2020clustering}. We define our validation sample using the spec--$z$'s that have not been used for training the photo--$z$'s. In Fig. \ref{fig:data2} we show results on the photo-$z$ validation sample. The left-hand side panel shows the bias $\Delta z=z_{\rm phot} - z_{\rm spec}$, the deviation of the photo-$z$ point estimate from the spec--$z$ measured for the same galaxy, as a function of spec--$z$. The bias is normalized by a $(1+z)$ factor, as this is often how this metric scales with redshift. The density plot is mostly flat around $\Delta z/(1+z)=0$ over the entire redshift range of interest, with a median value of -0.0014. Another photo-$z$ performance metric often used in photometric surveys is the scatter in the bias distribution, which can be presented as the normalized median absolute deviation (NMAD):
\begin{equation}
\sigma_{\rm NMAD} = 1.48 \times {\rm median}\Big( \frac{|\Delta z|}{1+z_{\rm spec}} \Big).    
\end{equation}
For the DESI Legacy Survey, we find $\sigma_{\rm NMAD}=0.01$. 

On the right hand side of Fig. \ref{fig:data2} we also show the mean bias and scatter in photo-$z$ bins of size $0.05$. The scatter is presented as $\sigma_{68}$, the 68th percentile width of the bias distribution about the median. Apart from the first redshift bin, which is mostly relevant only for one of the events we consider and is affected by a smaller validation sample, the bias shows an excellent photo-$z$ performance as it is always below 0.004. Also $\sigma_{68}$ is indicative of accurate redshifts, as it is below 0.02 over most of the redshift range. Using the scatter, we compute the $2\sigma$ and $3\sigma$ outlier fraction as the fraction of objects that lay more than 2 or 3 times $\sigma_{68}$ away from $\Delta z=0$. We find that the fractions are $\sim 4.5$ and $0.3\%$, respectively. These values are competitive with state-of-the-art cosmological photometric experiments such as DES. The DES science requirements include a maximum value for the scatter of $\sigma_{68}<0.12$, a requirement which is comfortably met by the sample described here. Both the scatter and bias are also competitive, with those found for the DES Science Verification and Year 1 photo-$z$ catalogs \citep{Sanchez_2014,Hoyle_2018} over a similar redshift range. 
The excellent results obtained can be understood in light of 1. the large overlap with spectroscopic surveys (larger than what is possible for DES) that provide an excellent training sample, 2. the fact that we are restricting the analysis to nearby redshifts, where the signal-to-noise of our photometry is best and more complete spectroscopic samples are available compared to larger redshifts, 3.  the lack of $i-$band photometry in the DESI imaging is not a determining factor in the quality of the photo-$z$'s since the 4000 \AA~ break is crosses to the $r$ band only at the edge of the redshift range considered, around 0.4. More information about the galaxies' Spectral Energy Distribution, other than the optical $grz$ data, is provided by the addition of the WISE data, which also contributes to the quality of the photo-$z$'s.

In order to ensure that our sample is volume-limited, we use an approach similar to that of \citet{palmese20_sts}. First, we define a maximum redshift of interest for each GW event, given its 90\% CI bounds in luminosity distance marginalized over the whole sky. The higher bound is converted into a maximum redshift given the largest value of $H_0$ we consider in the prior. For each event, given the maximum redshift considered, we derive the absolute magnitude that corresponds to the galaxy sample limiting apparent magnitude at that redshift. We then discard galaxies with an absolute magnitude below this threshold. Note that while we use a fiducial $\Lambda$CDM cosmology to derive these magnitudes, the $H_0$ dependence is irrelevant since the threshold value and the galaxies' absolute magnitudes scale with $H_0$ in the same way.

\begin{figure*}
\centering
\includegraphics[width=0.33\linewidth]{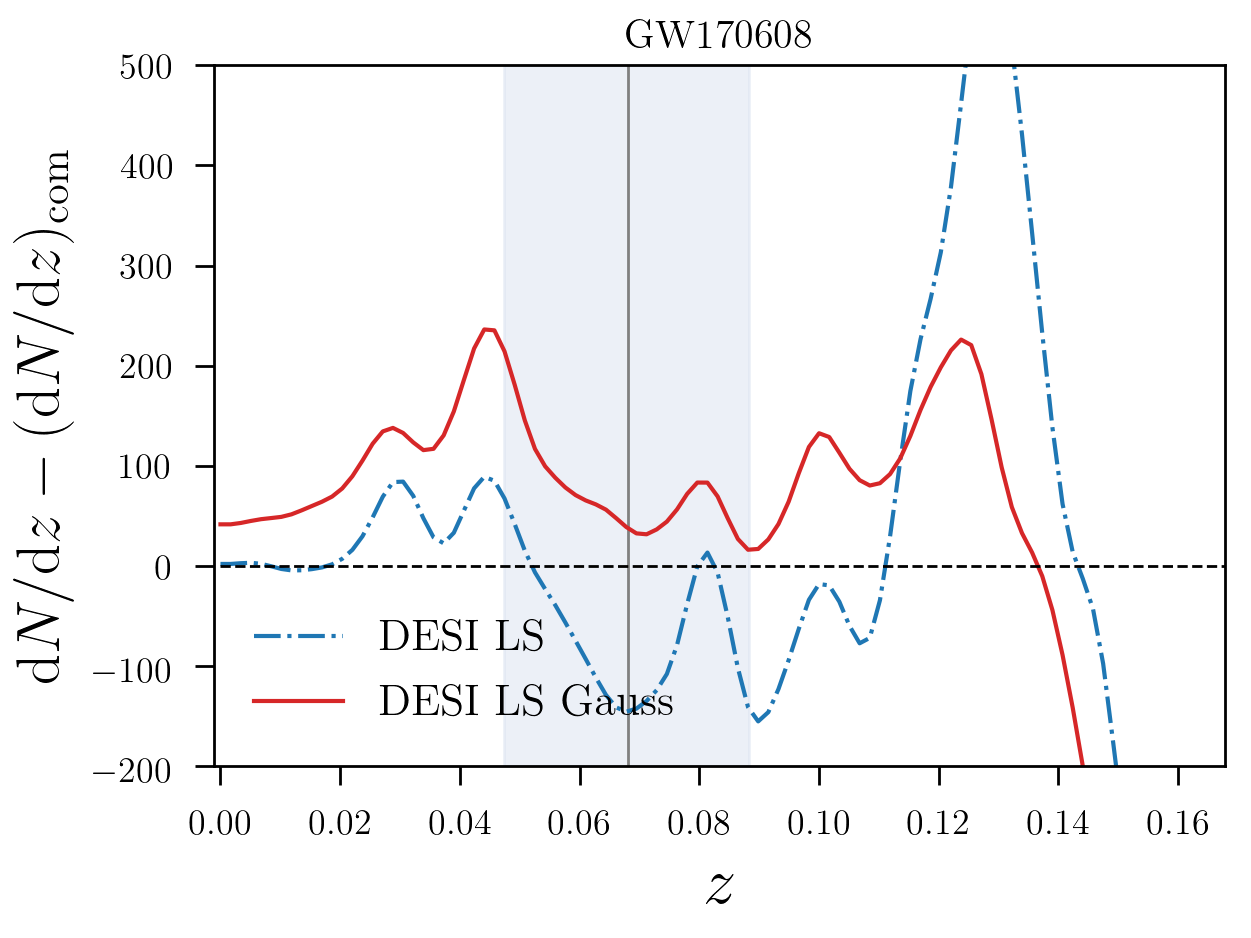}
\includegraphics[width=0.33\linewidth]{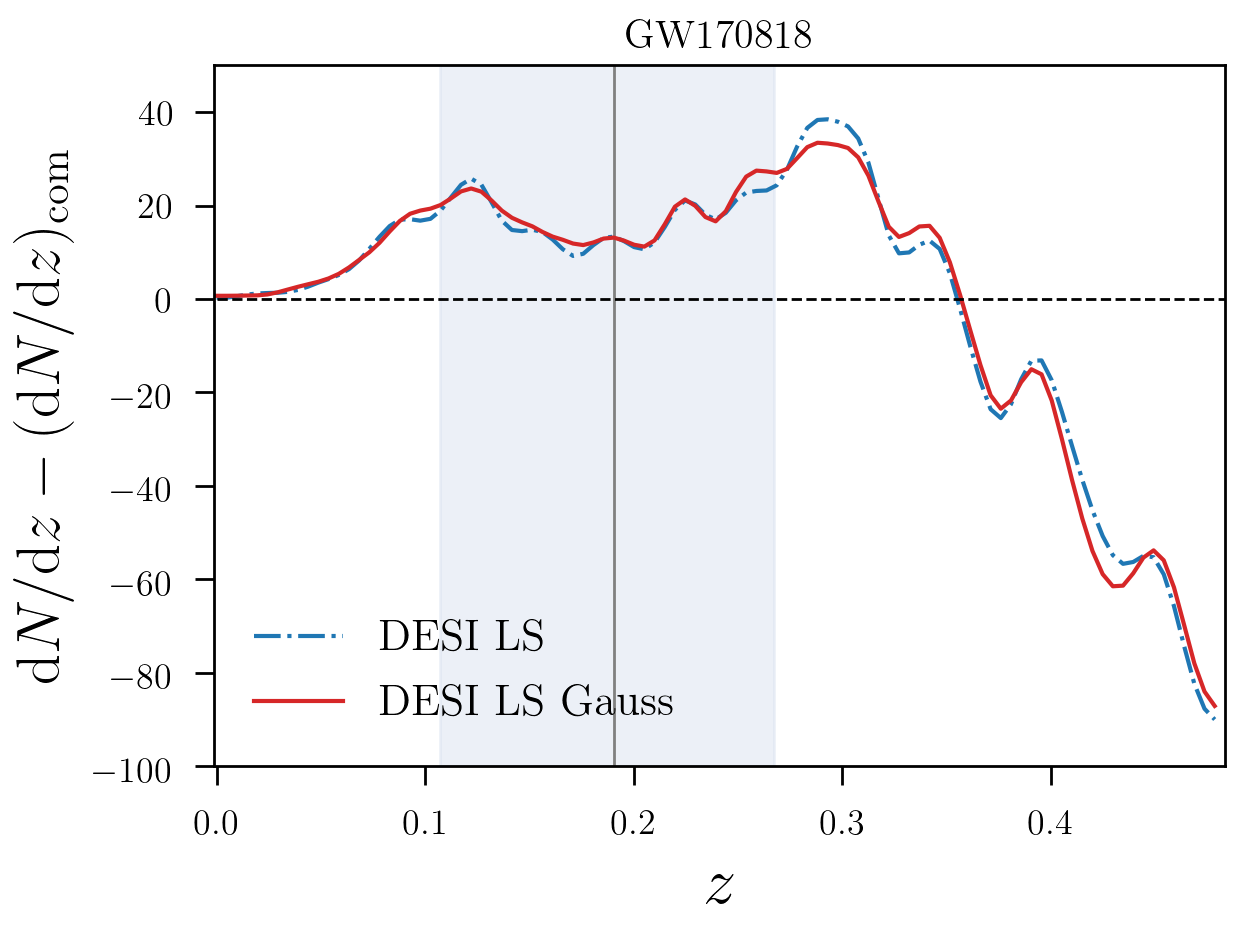}
\includegraphics[width=0.33\linewidth]{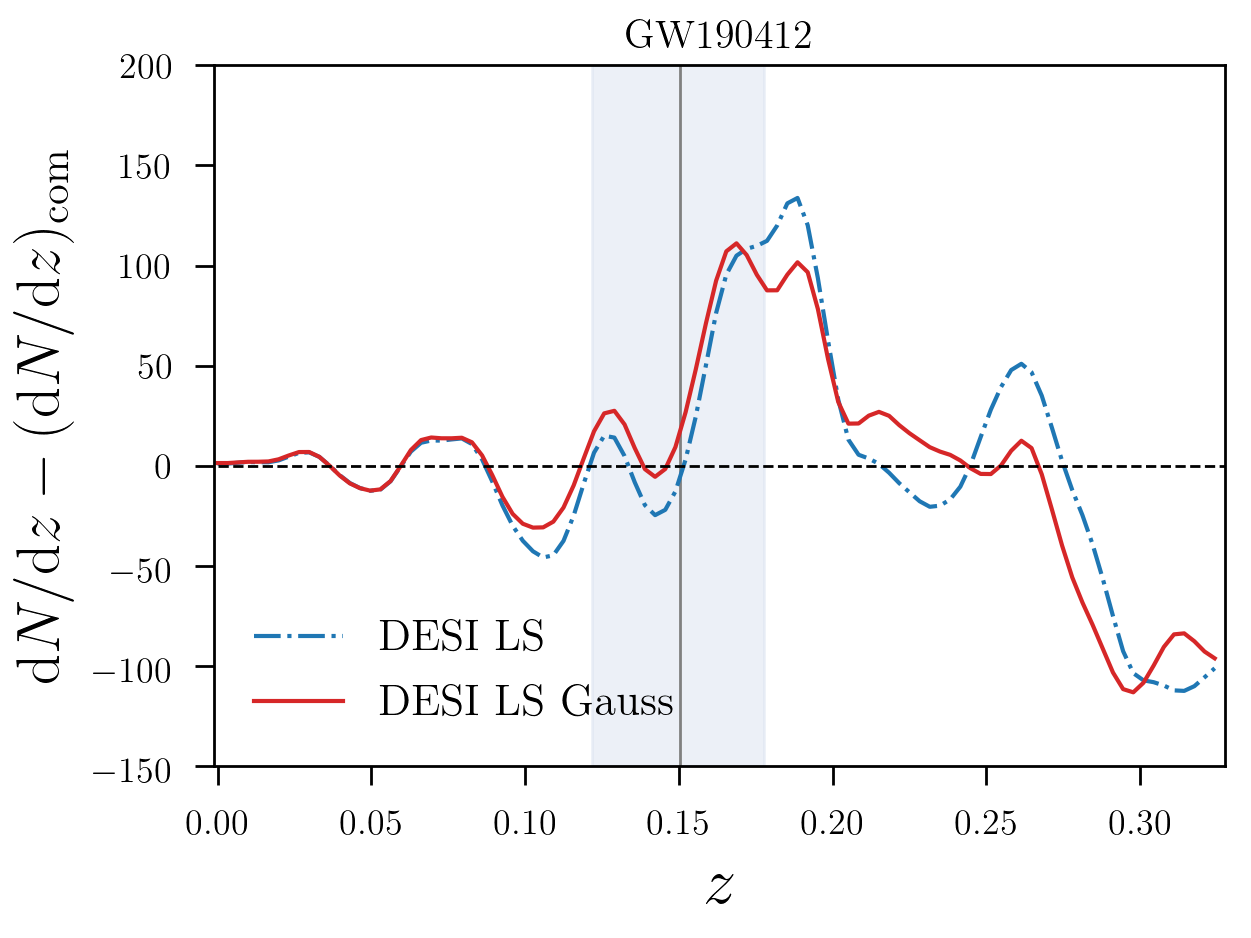}
\includegraphics[width=0.33\linewidth]{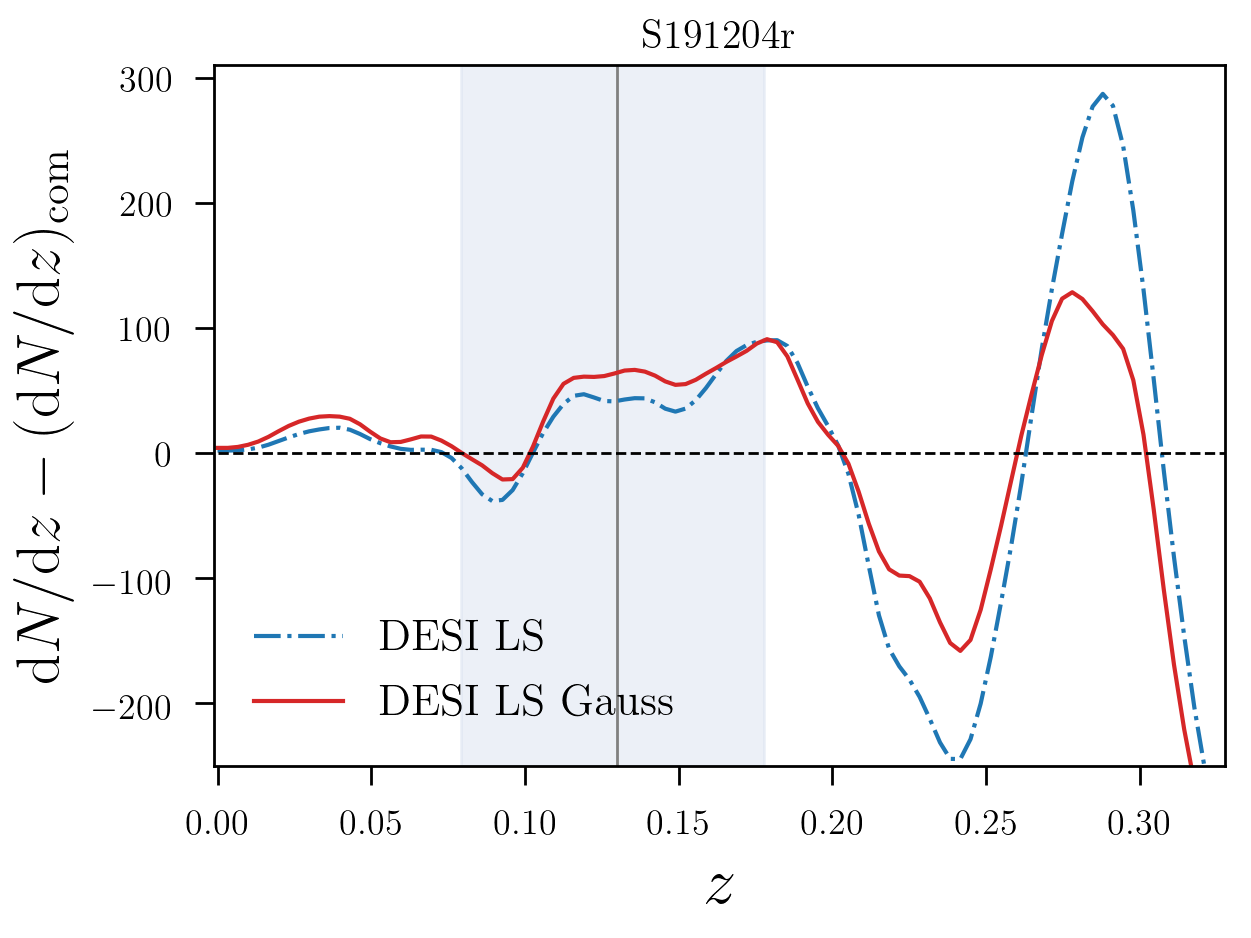}
\includegraphics[width=0.33\linewidth]{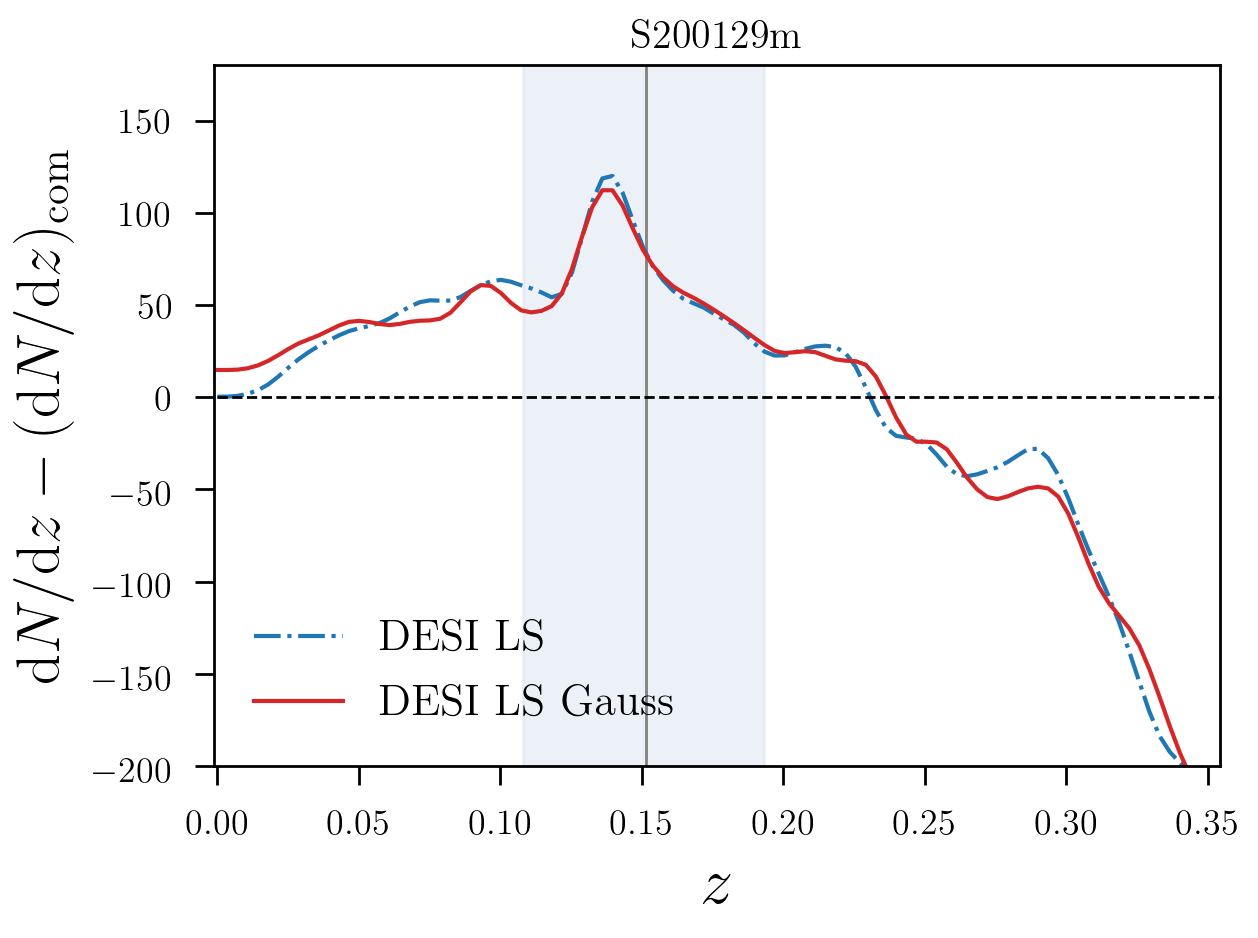}
\includegraphics[width=0.33\linewidth]{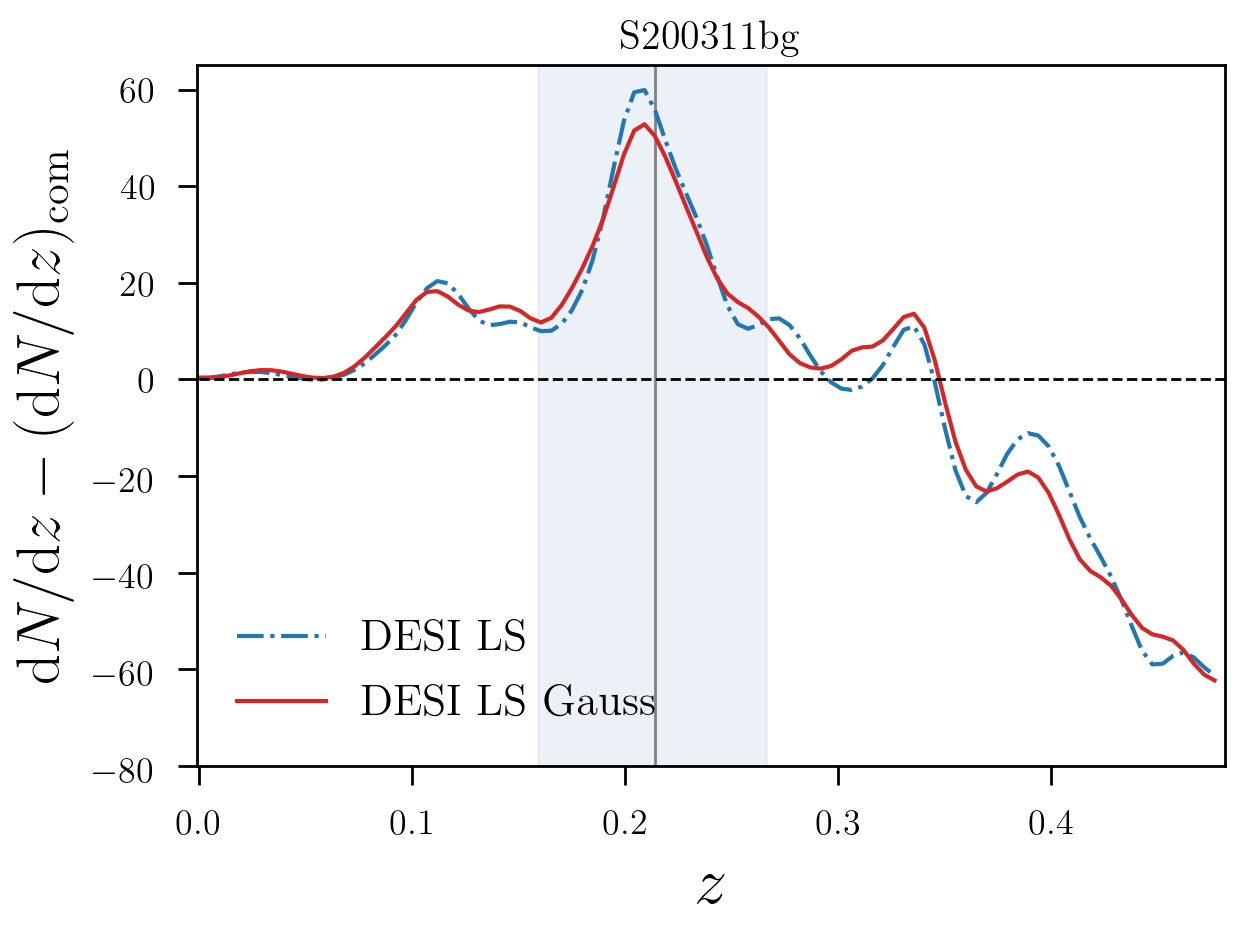}
\caption{Redshift distribution of galaxies in the 90\% CI area of the dark siren events analyzed in this work. The distribution is subtracted with a $dN/dz$ with uniform number density, in order to highlight the presence of overdensities and underdensities along the line of sight. The dashed blue line shows the distribution using the photometric redshifts point estimates from the DESI Legacy Survey presented in \citet{zhou2020clustering}, the dot-dash red line shows the same redshifts when their uncertainty is taken into account as a Gaussian error. The grey vertical lines represent the luminosity distance of each GW event marginalized over the entire sky, assuming an $H_0$ of 70 km s$^{-1}$ Mpc$^{-1}$, and the shaded regions are the $1\sigma$ uncertainties considering the same $H_0$. These regions are only showed for reference.}
\label{fig:dndz}
\end{figure*}

In Fig. \ref{fig:dndz} we show the redshift distribution of galaxies in the 90\% CI area of the GW events we analyze in this work. The distribution is subtracted with a $dN/dz$ that is uniform in comoving volume, in order to highlight the presence of overdensities and underdensities along the line of sight. The dot-dash blue line shows the distribution using the photometric redshifts point estimates from the DESI Legacy Survey described in this subsection, the solid red line shows the same redshifts when their uncertainty is taken into account as a Gaussian error. These two distributions are very close to each other, although the Gaussian sampling case tends, as expected, to smooth out some feature that are more prominent in the point estimate distribution. Remarkably, from these plots it is clear that, in most cases, the distance of the GW events, assuming an $H_0$ of 70 km s$^{-1}$ Mpc$^{-1}$ (given by the grey vertical line, with the shaded region reflecting the uncertainty in luminosity distance when marginalized over the entire sky), is close to overdensities in the galaxy distribution. These overdensities provide peaks in the $H_0$ posterior.

\subsubsection{Joint redshift-luminosity PDFs with Random Forests}

We compute joint redshift-luminosity PDFs using \texttt{GALPRO} \citep{galpro}, a \texttt{Python} package for generating multivariate PDFs of galaxy properties on-the-fly using Random Forests. We build a RF model to predict redshift and rest-frame $r$-band absolute magnitudes simultaneously. For this, we input the same features as used by \citet{zhou2020clustering} to predict photo-$z$'s. The only difference is that we convert fluxes into `asinh' magnitudes (`luptitudes') to avoid removing any faint galaxies with close to zero or negative fluxes in our target samples. The target features are the spec-zs and the rest-frame $r$-band absolute magnitudes we computed before. We train and test the model using the same samples used by the authors to train and test their RF model. We generate joint PDFs for galaxies in the test sample and use them to validate the model by checking probabilistic and marginal calibration. Finally, we use the trained model to generate joint PDFs for galaxies in our target datasets, for which we do not have the `truth' values.

To validate the marginal and joint PDFs of redshift and luminosity, we use the metrics described in \citet{galpro}. We check two different modes of calibration: probabilistic calibration and marginal calibration for the marginal PDFs, and their multivariate counterparts, probabilistic copula calibration and Kendall calibration respectively to validate the joint PDFs. We apply the metrics on PDFs of galaxies in the test dataset. 
The PIT distributions indicate that the recovered marginal PDFs are well-behaved, as they follow a nearly uniform distribution, with an outlier fraction below the percent level. The PIT results also show that the marginal PDFs are slightly overly broad and biased as the distributions have a convex shape and contain a slope. The marginal calibration plots make this concrete as the difference between the `true' and predicted cumulative distribution functions (CDFs) averages around zero, although with some fluctuations about the zero line. The copula probability integral transform (copPIT) distribution and the Kendall calibration are poor compared to their univariate counterparts. This indicates that the joint PDFs are not as accurate as the marginal ones. However, we will only be using the marginal PDFs in this work. 
To improve the calibration of the PDFs, we try different methods. At first, we tune the model by incrementing the minimum number of samples (\texttt{min\_samples\_leaf}) that must be present in a leaf node from $1$ (the default) to $2$, $3$, and $5$. However, this results in further degradation of the calibration. Next, we try augmenting the training dataset by scattering the galaxies. Specifically, we do this by scattering the photometry according to the photometric errors. Each galaxy in the training dataset is scattered five times, and the galaxies in the testing dataset are scattered once for consistency. We run the analysis again and find that there is a slight improvement in the PIT distributions as they become more uniform, so we choose this training for our analysis. 
Each galaxy is taken into account in the Bayesian framework detailed in \S\ref{method} with 100 photo-$z$ and luminosity samples. We consider the samples that pass the luminosity cut as described in \S\ref{sec:LSphz}, and we use the kernel density estimation (KDE) method to retrieve the redshift PDFs for the galaxies in our sample.

\subsubsection{Galaxy fakes}

Two of the events considered, S200129m and GW170608, are covered to less than 90\% CI in area by the DESI imaging. For the uncovered regions of the 90\% CI, we inject galaxy fakes that are samples from our prior distribution, in order to ensure that the marginalization occurs over all the possible host galaxies and that the final uncertainty on the Hubble constant is not underestimated. The prior distribution of galaxies is given in both cases considered here, the DESI imaging photo--$z$'s and the GALPRO PDFs, by the training sample. We take Monte Carlo samples from the redshift/luminosity distribution to assign to each fake a redshift and a luminosity The spatial distribution is uniform per unit solid angle, with a number density of galaxies provided by the mean value of the training sample. We repeat the redshift/luminosity sampling 100 times per galaxy for the case of the GALPRO measurements, in order to match the number of PDF samples considered for the real galaxies.

\section{Method}\label{method}

The formalism used in this work is adapted from \citet{chen17}, following \citet*{darksiren1} and \citet{palmese20_sts}. The $H_0$ posterior derivation is described in detail in those works. Here we highlight that the posterior is computed within a Bayesian formalism. In the first step, we write the  posterior probability of $H_0$ as it follows from Bayes' theorem:
\begin{equation}
p(H_0|d_{\rm GW},d_{\rm EM})  \propto p(d_{\rm GW},d_{\rm EM}|H_0)p(H_0)\, ,
\label{eq:posteriormain}
\end{equation}
where $d_{\rm GW}$ is the GW data from a single event, $d_{\rm EM}$ is the data from the galaxy survey into consideration, $p(d_{\rm GW},d_{\rm EM}|H_0)$ is the joint GW--EM likelihood, and $p(H_0)$ is the prior on $H_0$ which we take to be uniform over the range $[20,140]$ km s$^{-1}$ Mpc$^{-1}$. After marginalizing over all potential host galaxies $i$ in the galaxy sample, over the unknown true redshift $z$ and over the photometric redshift bias $\Delta z$, one can show that Eq. (\ref{eq:posteriormain}) can be rewritten as:
\begin{widetext}
\begin{equation}
p(H_0|d_{\rm GW}, d_{\rm EM}) \propto \frac{p(H_0)}{\beta (H_0)} \sum_i \frac{1}{\mathcal{Z}_i}\int \de z ~\de \Delta z \, p(d_{\rm GW}|d_L(z,H_0),\hat{\Omega}_i) p_i(d_{\rm EM}|z,\Delta z) ~p(\Delta z) \frac{r^2(z)}{H(z)} \, .\label{eq:like3}
\end{equation}
\end{widetext}
where $\beta (H_0)$ is the likelihood normalization factor that takes into account selection effects, and $\mathcal{Z}_i = \int p(d_{\rm EM} | z_i) r^2(z_i)/H(z_i) ~\de z_i$ are evidence terms that normalize the posterior. The first term in the integral is the GW likelihood computed at the position and redshift of the galaxy $i$, the second term is the redshift likelihood of the galaxy, shifted by the photo--$z$ bias, and $p(\Delta z)$ is the prior on the photo--$z$ bias, which we measure from the redshift validation sample.

Selection effects are taken into account in the $\beta (H_0)$ term as follows, similarly to what is done in \citet{chen17} and \citet{gray2019cosmological}. This term is the joint GW-EM likelihood marginalized over all possible GW and EM data.
We consider the DESI legacy survey complete down to the absolute magnitudes considered for the events taken into account, for which we only consider galaxies out to $z<0.5$. 
Under this assumption and that of isotropy of the events on large scales, after marginalizing over the entire sky, the normalization factor reduces to:
\begin{equation}
    \beta(H_0)=\int p_{\rm sel}^{\rm GW} (d_L(z,H_0)) p(z) \de z
\end{equation}
where $p_{\rm sel}^{\rm GW}$ describes our selection function of GW events, and it is 1 if the detector network SNR is $>12$ and the localization volume satisfies our selection criteria, 0 otherwise. Note this is the same as Eq. 15 in \citet{chen17}, where a full derivation can be found. 

In order to compute $\beta(H_0)$, we adopt a Monte Carlo approach. We simulate 70,000 BBH mergers and compute $\beta (H_0)$ for 20 values of $H_0$ within our prior range. We generate GW events using the \texttt{BAYESTAR} software \citep{bayestar,Singer_2016,Singer_supp}, also based on tools from \texttt{LALSuite} \citep{lalsuite}. We assume the O3 sensitivity curves for LIGO and Virgo measured from the first months of O3 operations, as published by LIGO in Document P1200087 (\url{https://dcc.ligo.org/LIGO-P1200087/public}). The injected events follow a distribution that is uniform in comoving volume, assuming a Planck 2018 cosmology with different $H_0$ values. We assume IMRPhenomD waveforms both for the injections and reconstructions.
The BHs follow a mass function described by a power-law with index 1.6 (as found in \citealt{Properties_O3a} for the ``power law + peak'' mass function), and a uniform spin distribution between $(-1,1)$. After the 70,000 injections are made, we run a matched-filter search to retrieve the detected events. A detection is made when at least 2 detectors reach a single--detector signal--to--noise ratio SNR$>4$ and the network SNR is $>12$. Gaussian noise is added to the measurement. In the last step, we reconstruct \texttt{BAYESTAR} skymaps for the detection. The reconstruction is made assuming a distance prior which scales as $\propto d_L^2$, where $d_L$ is the luminosity distance. 

Another component of our selection is the requirement that the high probability region of the GW map is well-covered by the DESI imaging. Because we do not expect the DESI footprint to be correlated with the GW antenna pattern, we do not need to take into account this selection effect on the GW events, as if we were performing an isotropic random sampling of the events. We find that this selection results in a $\beta(H_0)$ that follows closely the $H_0^{3}$ function used in previous works.

We note that the selection effects calculation could be improved by considering the same priors as those assumed in the GW likelihood. In other words, it could be improved by assuming the same priors used when determining the sky map. The most significant difference in the priors concerns the binary mass. This is assumed to be uniform on the redshifted detector-frame mass when computing the sky map, while the priors we have considered for the selection function calculation assume an astrophysical mass distribution on the source frame mass. However, \citet{gray2019cosmological} show that if the mass distribution takes the form of a power law, no prior correction is required for this difference in assumptions. In the future, one may prefer to assume a more realistic mass function, and ideally simultaneously fit for both the mass function and the cosmology.

Assuming that the data $\{d_{{\rm GW},i}\}$ from a sample of GW events is made of events that are independent of each other, the Hubble constant posterior can easily be rewritten for a combination of GW events through the product of the single event $j$ likelihoods:
\begin{equation}
p(H_0|\{d_{{\rm GW},i}\},d_{{\rm EM}}) \propto p(H_0) \prod_j p(d_{{\rm GW},j},d_{{\rm EM}}|H_0)\, .
\end{equation}

\section{Results and Discussion} \label{results}

\begin{figure*}
    \centering
    \includegraphics[width=0.8\linewidth]{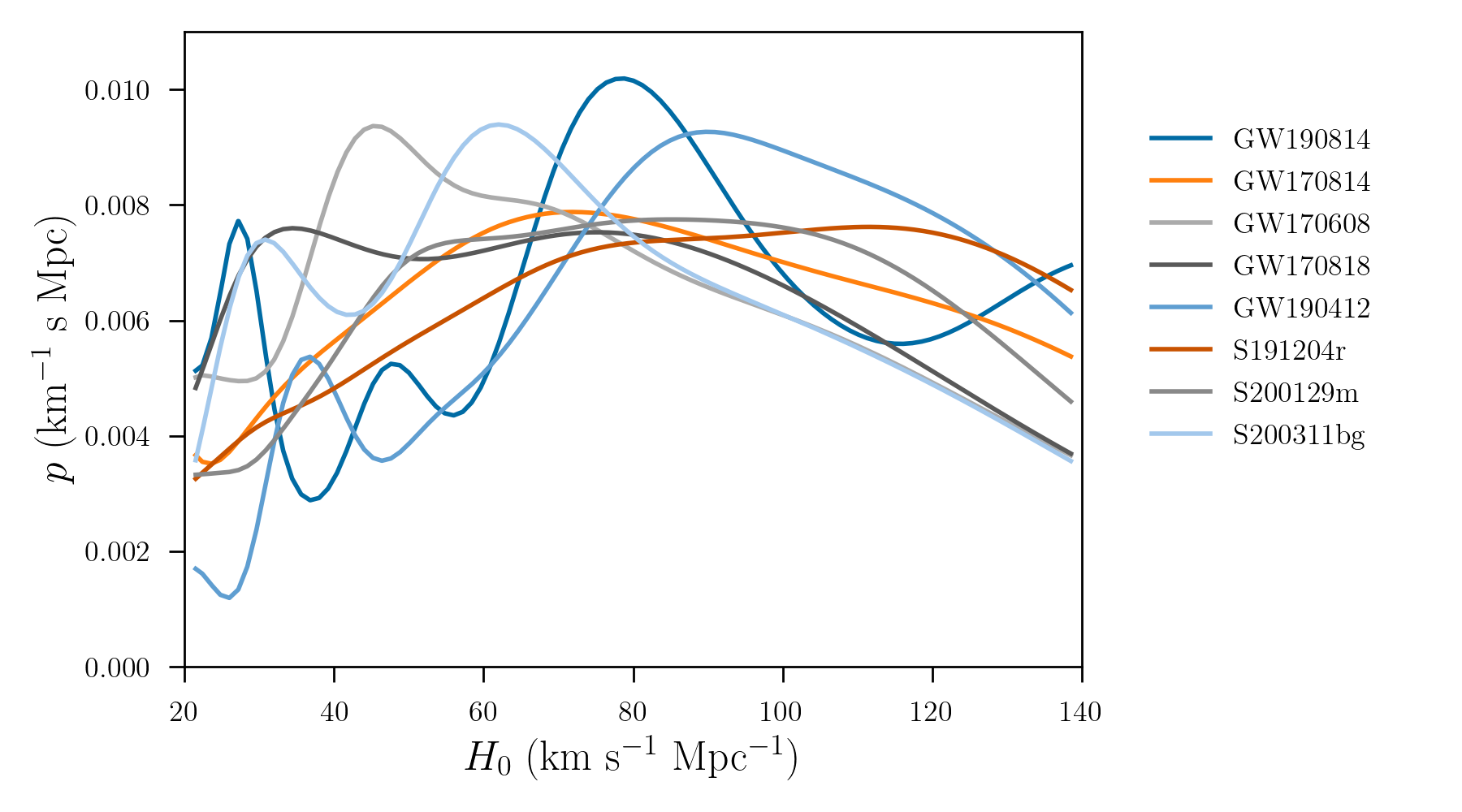}
    \caption{Hubble constant posterior distributions for the dark sirens considered in this work. Each line indicates the posterior from a single GW event and galaxies from DES or the DESI Imaging Survey.}
    \label{fig:posteriors}
\end{figure*}

\begin{figure*}
    \centering
    \includegraphics[width=0.8\linewidth]{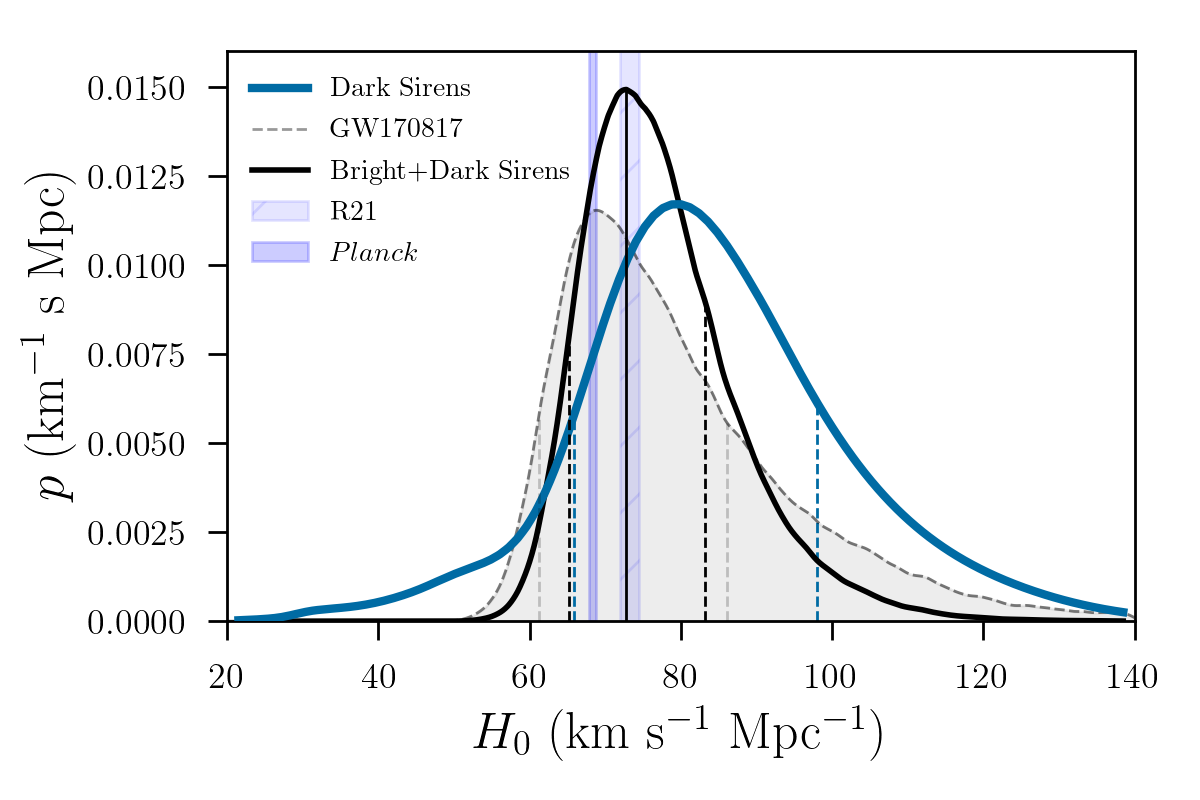}
    \caption{Hubble constant posterior distributions. The blue line shows the result from the combination of all dark sirens considered in this paper. The shaded grey posterior represents the GW170817 standard siren result adapted from \citet{nicolaou2019impact}, which makes use of the presence of the electromagnetic counterpart. The black posterior represents the final result of this work, showing the joint constraint from both the bright (i.e. GW170817) and the dark standard sirens. The vertical dashed lines show the 68\% region for each posterior. For reference, the $1\sigma$ \citet{planck18} and \citet{riess2021} (R21) constraints on $H_0$ are also shown as the vertical shaded regions. Posteriors are arbitrarily rescaled only for visualization purposes.}
    \label{fig:result}
\end{figure*}

In this Section, we show the results using the DESI Legacy Survey photo--$z$'s, replaced by the spectroscopic redshifts described in \S\ref{sec:spec}, where available. We produce an $H_0$ posterior for each of the events GW170608, GW170818, GW190412, S191204r, S200129m and S200311bg, and use the DES posteriors published in \citet{palmese20_sts} for GW170814 and GW190814.
As reported in Section \ref{data}, the metrics found for the GALPRO results are not as optimal as those we find for the Legacy Survey photo-$z$'s, but we still use them as an alternative photo-$z$ method for comparison, and find that the $H_0$ constraints do not significantly depend on which of the photo--$z$'s are used. We provide the Legacy Survey photo-$z$ results as our fiducial result.

The posterior distributions on $H_0$ for the single dark siren events considered (in colors) are shown in Fig. \ref{fig:posteriors}. Each event reduces the 68\% CI of the $H_0$ prior to its $\sim 85\%$. The most constraining event is GW190412, which is reasonable provided that this event has the best localization after GW190814, and its $1\sigma$ constraint reaches close to 80\% of the $H_0$ prior. Fig. \ref{fig:dndz} also shows a prominent overdensity of galaxies around redshift 0.19 for GW190412, which is reflected in the $H_0$ posterior as a peak around $H_0$ of 80 km s$^{-1}$ Mpc$^{-1}$. We stress that it is expected that the $H_0$ posteriors from individual dark standard sirens present multiple peaks, as these correspond to multiple overdensities in redshift along the line of sight. It is also expected that the peaks are very broad for the current level of precision: the GW localization contains thousands of potential host galaxies, over which we marginalize, and both the GW events luminosity distance estimate and the photometric redshifts are measured to a precision of 10-30\%.  We find that for both events where fakes are required, the effect of their addition is that of further flattening the posterior compared to the case with no fakes. This is expected since it effectively corresponds to adding more galaxies for marginalization, and with an uninformative redshift distribution. It is after combining a large enough number of events (namely $\mathcal{O}(100)$ for a $\mathcal{O}(1\%)$ precision on $H_0$; \citealt{palmese20_sts}) that the dark siren method becomes more powerful.
Fig. \ref{fig:result} shows the posterior obtained by combining all the dark sirens in blue. The maximum \emph{a posteriori} with the 68\% CI is $79.8^{+19.1}_{-12.8}$ km s$^{-1}$ Mpc$^{-1}$ from this combination. This result is broadly consistent within $1\sigma$ with both the \emph{Planck} and \citet{riess2021} constraints, whose 68\% intervals are represented for reference by the vertical shaded regions in Fig. \ref{fig:result}.

\citet{gwtc2} present results on GW190412 that assume two different sets of fiducial models, namely the SEOBNRv4PHM and a set of empirical models called IMRPhenomPv3HM. We have tested that the choice of the map does not result in any significant effect on the $H_0$ posterior distribution, and choose to use the former version.

The origin of the BBH mergers that LIGO/Virgo have detected is unclear, and it is likely that multiple formation mechanisms are at play \citep{zevin2021}. While it is plausible that some channels may occur preferentially in low-mass galaxies (e.g. \citealt{palmese_conselice}), an assumption that several dark siren works have made is that higher mass, more luminous galaxies are more likely to host BBH events. Similarly, we weigh the galaxies in our sample based on their $r$-band luminosity. Since this approach effectively gives more weight to the less numerous, more luminous galaxies, compared to the lower mass galaxies, we expect a luminosity weighting to produce more stringent constraints. We find that the weighting produces a slight improvement on the constraints, yielding a 1\% improvement on the precision. We prefer to provide the more conservative result that does not include the luminosity weighting.

We note that the results presented here are valid under a precise cosmological model, the Flat $\Lambda$CDM model. This is unlike the $H_0$ measurement from GW170817 \citet{2017Natur.551...85A}, which is nearby enough to only be sensitive to the Hubble constant. The measurement presented in this work, similarly to all other BBH dark siren measurements available so far, depend on the background cosmology because they extend beyond redshift $\sim 0.1$, where the Hubble parameter becomes more sensitive to the matter density $\Omega_m$ and other parameters. We have tested that our results are not significantly affected by changes to $\Omega_m$ within the $2\sigma$ interval found by the DES-only measurements presented in \citet{DESY1_combined}, so that our $H_0$ constraint, although tied to a Flat $\Lambda CDM$ model, is not dependent on the exact $\Omega_m$ value assumed, at least within a prior consistent with this DES constraints. Future dark siren measurements from a larger sample of events with more precise distance measurements will be sensitive to the matter density and other parameters, and should be able to place constraints on $\Omega_m$ and the dark energy equation of state (e.g. \citealt{laghi}).

We compare our results to that of \citet{finke2021cosmology}, who find $H_0=67.3^{+27.6}_{-17.9}$ km s$^{-1}$ Mpc$^{-1}$ from a combination of a large number of events from the O1, O2, and the first half of O3. Our result is broadly consistent with it, but we find a more precise constraint despite the smaller number of events considered. The galaxy catalog used in that work, the GLADE catalog \citet{glade}, is a compilation of galaxies from different surveys complete out to a distance of $\sim 37$ Mpc \citep{glade}, which is extremely useful for targeted follow-up of nearby GW events. However, at the larger distances of the events considered in \citet{finke2021cosmology} and this work, GLADE is highly incomplete and inhomogeneous, as opposed to the DESI imaging. 
Only 5 events in \citet{finke2021cosmology} are considered to be largely covered by the GLADE catalog, and are therefore expected to provide a useful contribution to the final $H_0$ constraint. However, one of those, GW150914, is at a problematic location for galaxy catalogs, as a large portion of its 90\% CI area is either covered by the Large Magellanic Cloud, or close to the Galactic Plane. Most of the contribution to their $H_0$ posterior comes from GW190814, while they properly take into account the catalog incompleteness for the remaining events, which has to result in less constraining results as discussed above. For these reasons, and thanks to a careful choice of the events to be considered (i.e. those that are better localized), we are able to retrieve a more constraining Hubble constant posterior despite the use of a smaller number of events. We also note that our $H_0$ prior is slightly larger than theirs ([30,140] km s$^{-1}$ Mpc$^{-1}$), which could be misleading when comparing the precision of measurements.
 A preprint showing contsraints from a larger sample of BBH became available at the time this work was being submitted \citep{theligoscientificcollaboration2021constraints}, so we do not provide an extensive comparison. We only note that \citet{theligoscientificcollaboration2021constraints} use a galaxy catalog that is very similar to that in \citet{finke2021cosmology}, and will therefore suffer from similar limitations from the host galaxies standpoint.

At last, we combine our results with those from GW170817 assuming the electromagnetic counterpart. We use the posterior from \citet{nicolaou2019impact}, who more carefully revisit the peculiar velocity contribution to the constrain compared to the original measurement of \citet{2017Natur.551...85A}, with the $H_0$ prior appropriately rescaled to match the one we use for the dark sirens. The posterior obtained by the combination of all dark standard sirens considered in this work, and GW170817, is shown by the black line in Fig. \ref{fig:result}, and it is also broadly consistent with \citet{planck18} and \citet{riess2021}. Our final $H_0$ constraint from this analysis is $72.77^{+11.0}_{-7.55}$ km s$^{-1}$ Mpc$^{-1}$. The constraint is largely driven by the one bright standard siren available, but the dark standard sirens also provide a significant contribution by improving the precision from GW170817 by 28\%.

\begin{table*}
\centering
\begin{tabular}{cccccc}
Event(s) & Prior  & $H_0 $ & $\sigma_{H_0}/H_0$ & $\sigma_{H_0}/\sigma_{\rm prior}$ & Reference\\
\hline
GW170817 - bright & $[20,140]$ & $68.80 ^{+ 17.3}_{-7.63}$ & 18\% & 31\% & Adapted from \citet{nicolaou2019impact}\\
F21 O1-O3a -- dark & $[30,140]$ & $67.3^{+ 27.6}_{-17.9}$ & 34\% & 61\% & \citet{finke2021cosmology}\\
O1-O3 -- dark & $[20,140]$ & $79.8^{+19.1}_{-12.8}$ & 20\% & 39\% & This work\\
O1-O3 -- all & $[20,140]$ & $72.77^{+11.0}_{-7.55}$ & 12\% & 22\% & This work\\
\end{tabular}
\caption{Hubble constant measurements using gravitational wave standard sirens from this work and previous works. $H_0$ values and uncertainties are given in $~{\rm km~s^{-1}~Mpc^{-1}}$, and $H_0$ priors are flat. The uncertainty from the flat prior only is derived by assuming the same $H_0$ maximum found in the analysis. Quoted uncertainties represent 68\% HDI around the maximum of the posterior. The ``$\sigma_{H_0}/\sigma_{\rm prior}$'' column shows the 68\% CI from the posterior divided by 68\% CI of the
prior width. } 
\label{tab:results}
\end{table*}

\section{Conclusions}\label{conclusions}

In this paper, we have presented a new measurement of the Hubble constant using the best available gravitational wave events up to date and a state-of-the-art, uniform galaxy catalog from the DESI Legacy Survey. This measurement results in a constraint of the Hubble constant of $79.8^{+19.1}_{-12.8}$ km s$^{-1}$ Mpc$^{-1}$, i.e. a $\sim 20\%$ uncertainty on $H_0$ from dark sirens alone. This is the most constraining dark standard siren measurement obtained so far using a complete galaxy catalog. After combination with the one bright standard siren available, properly marginalized over different models for the peculiar velocity of the host galaxy, we obtain $H_0=72.77^{+11.0}_{-7.55}$ km s$^{-1}$ Mpc$^{-1}$, a $12\%$ measurement of $H_0$. We note that the combination of the GW170817 bright siren, whose $H_0$ estimate is independent of the cosmological model, with the dark sirens, which we derived within the assumption of a Flat $\Lambda$CDM scenario, is also tied to a Flat $\Lambda$CDM scenario.

In the near future, spectroscopic observations of the host galaxies in the localization regions, higher order multipole analyses of the GW candidates, and deep observations of other well-localized events not covered by the DESI Legacy Survey (e.g. GW150914 and S191216ap) could further improve the measurement presented in this work. However, the significant improvement on dark siren measurements will be possible with the upcoming LIGO/Virgo/KAGRA observing run, expected to start in the second half of 2022. With the dark siren precision expected to be achieved after a few years of LIGO/Virgo/KAGRA running at design sensitivity or better (when a 2\% statistical precision on $H_0$ will become possible, and such precision is valuable to inform us on the Hubble constant tension), careful studies of potential systematics not included here should be carried out, especially in light of new results from population studies of BBH: e.g. the impact of a galaxy catalog depth on the constraints, based on different BBH formation channels; the impact of the Gaussian ansatz on the dark siren posterior.

\acknowledgments
\noindent AP thanks James Annis, Zoheyr Doctor, Will Farr, Maya Fishbach, Daniel Holz, Ignacio Maga\~na-Hernandez, and Marcelle Soares-Santos for very useful help and discussion on standard sirens. AP also thanks Segev BenZvi, ChangHoon Hahn, and Rongpu Zhou for help with the DESI Legacy Survey, and Constantina Nicolaou for sharing the GW170817 posterior. Support for this work was provided by NASA through the NASA Hubble Fellowship grant HST-HF2-51488.001-A awarded by the Space Telescope Science Institute, which is operated by Association of Universities for Research in Astronomy, Inc., for NASA, under contract NAS5-26555. In this work we made extensive use of TOPCAT \citep{topcat}.

\noindent The Legacy Surveys consist of three individual and complementary projects: the Dark Energy Camera Legacy Survey (DECaLS; NSF's OIR Lab Proposal ID 2014B-0404; PIs: David Schlegel and Arjun Dey), the Beijing-Arizona Sky Survey (BASS; NSF's OIR Lab Proposal ID 2015A-0801; PIs: Zhou Xu and Xiaohui Fan), and the Mayall z-band Legacy Survey (MzLS; NSF's OIR Lab Proposal ID 2016A-0453; PI: Arjun Dey). DECaLS, BASS and MzLS together include data obtained, respectively, at the Blanco telescope, Cerro Tololo Inter-American Observatory, The NSF's National Optical-Infrared Astronomy Research Laboratory (NSF's OIR Lab); the Bok telescope, Steward Observatory, University of Arizona; and the Mayall telescope, Kitt Peak National Observatory, NSF's OIR Lab. The Legacy Surveys project is honored to be permitted to conduct astronomical research on Iolkam Du'ag (Kitt Peak), a mountain with particular significance to the Tohono O'odham Nation.

\noindent The NSF's OIR Lab is operated by the Association of Universities for Research in Astronomy (AURA) under a cooperative agreement with the National Science Foundation.

\noindent This project used data obtained with the Dark Energy Camera (DECam), which was constructed by the Dark Energy Survey (DES) collaboration. Funding for the DES Projects has been provided by the U.S. Department of Energy, the U.S. National Science Foundation, the Ministry of Science and Education of Spain, the Science and Technology Facilities Council of the United Kingdom, the Higher Education Funding Council for England, the National Center for Supercomputing Applications at the University of Illinois at Urbana-Champaign, the Kavli Institute of Cosmological Physics at the University of Chicago, Center for Cosmology and Astro-Particle Physics at the Ohio State University, the Mitchell Institute for Fundamental Physics and Astronomy at Texas A\&M University, Financiadora de Estudos e Projetos, Fundacao Carlos Chagas Filho de Amparo, Financiadora de Estudos e Projetos, Fundacao Carlos Chagas Filho de Amparo a Pesquisa do Estado do Rio de Janeiro, Conselho Nacional de Desenvolvimento Cientifico e Tecnologico and the Ministerio da Ciencia, Tecnologia e Inovacao, the Deutsche Forschungsgemeinschaft and the Collaborating Institutions in the Dark Energy Survey. The Collaborating Institutions are Argonne National Laboratory, the University of California at Santa Cruz, the University of Cambridge, Centro de Investigaciones Energeticas, Medioambientales y Tecnologicas-Madrid, the University of Chicago, University College London, the DES-Brazil Consortium, the University of Edinburgh, the Eidgenossische Technische Hochschule (ETH) Zurich, Fermi National Accelerator Laboratory, the University of Illinois at Urbana-Champaign, the Institut de Ciencies de l'Espai (IEEC/CSIC), the Institut de Fisica d'Altes Energies, Lawrence Berkeley National Laboratory, the Ludwig-Maximilians Universitat Munchen and the associated Excellence Cluster Universe, the University of Michigan, the National Optical Astronomy Observatory, the University of Nottingham, the Ohio State University, the University of Pennsylvania, the University of Portsmouth, SLAC National Accelerator Laboratory, Stanford University, the University of Sussex, and Texas A\&M University.

\noindent BASS is a key project of the Telescope Access Program (TAP), which has been funded by the National Astronomical Observatories of China, the Chinese Academy of Sciences (the Strategic Priority Research Program "The Emergence of Cosmological Structures" Grant \# XDB09000000), and the Special Fund for Astronomy from the Ministry of Finance. The BASS is also supported by the External Cooperation Program of Chinese Academy of Sciences (Grant \# 114A11KYSB20160057), and Chinese National Natural Science Foundation (Grant \# 11433005).

\noindent The Legacy Survey team makes use of data products from the Near-Earth Object Wide-field Infrared Survey Explorer (NEOWISE), which is a project of the Jet Propulsion Laboratory/California Institute of Technology. NEOWISE is funded by the National Aeronautics and Space Administration.

\noindent The Legacy Surveys imaging of the DESI footprint is supported by the Director, Office of Science, Office of High Energy Physics of the U.S. Department of Energy under Contract No. DE-AC02-05CH1123, by the National Energy Research Scientific Computing Center, a DOE Office of Science User Facility under the same contract; and by the U.S. National Science Foundation, Division of Astronomical Sciences under Contract No. AST-0950945 to NOAO.

\noindent The Photometric Redshifts for the Legacy Surveys (PRLS) catalog used in this paper was produced thanks to funding from the U.S. Department of Energy Office of Science, Office of High Energy Physics via grant DE-SC0007914.

\bibliographystyle{yahapj_twoauthor_arxiv_amp}
\bibliography{references}

\end{document}